\documentclass[12pt]{article}
\usepackage{graphicx}
\usepackage{color}
\usepackage{multirow}

\newcommand\pubnumber{NuPhys2016-Yokoyama}
\newcommand\pubdate{April 2017}

\def\utokyo{Department of Physics, Graduate School of Science\\
The University of Tokyo, 113-0033, Tokyo, JAPAN}

\def\Title#1{\begin{center} {\Large #1 } \end{center}}
\def\Author#1{\begin{center}{ \sc #1} \end{center}}
\def\Address#1{\begin{center}{ \it #1} \end{center}}

\newcommand\pubblock{\rightline{\begin{tabular}{l} \pubnumber\\
         \pubdate  \end{tabular}}}
\newenvironment{Abstract}{\begin{quotation}  }{\end{quotation}}
\newenvironment{Presented}{\begin{quotation} \begin{center} 
             PRESENTED AT\end{center}\bigskip 
      \begin{center}\begin{large}}{\end{large}\end{center} \end{quotation}}
\def\Acknowledgements{\bigskip  \bigskip \begin{center} \begin{large}
             \bf ACKNOWLEDGEMENTS \end{large}\end{center}}

\def\beq{\begin{equation}}
\def\eeq#1{\label{#1}\end{equation}}
\def\eeqn{\end{equation}}

\def\beqa{\begin{eqnarray}}
\def\eeqa#1{\label{#1}\end{eqnarray}}
\def\eeqan{\end{eqnarray}}

\let\bar=\overbar

\def\Dslash{\not{\hbox{\kern-4pt $D$}}}
\def\dslash{\not{\hbox{\kern-2pt $\del$}}}

\def\msb{{\bar{\ssstyle M \kern -1pt S}}}

\begin{document}
\begin{titlepage}
\pubblock

\vfill
\Title{The Hyper-Kamiokande Experiment}
\vfill
\Author{ Masashi Yokoyama\\ For The Hyper-Kamiokande Proto-Collaboration }
\Address{\utokyo}
\vfill
\begin{Abstract}
Hyper-Kamiokande (HK) is a next generation large water Cherenkov detector to be built in Japan,
based on the highly successful Super-Kamiokande detector.
HK will offer a broad science program such as neutrino oscillation studies, proton decay searches, and neutrino astrophysics with unprecedented sensitivities.
This paper describes the overview and physics potential of HK.
\end{Abstract}
\vfill
\begin{Presented}
NuPhys2016, Prospects in Neutrino Physics \\
Barbican Centre, London, UK,  December 12--14, 2016
\end{Presented}
\vfill
\end{titlepage}
\def\thefootnote{\fnsymbol{footnote}}
\setcounter{footnote}{0}

\newcommand{\numu}{\ensuremath{\nu_{\mu}}}                   
\newcommand{\numubar}{\ensuremath{\overline{\nu}_{\mu}}}                   
\newcommand{\nue}{\ensuremath{\nu_{e}}}                   
\newcommand{\nuebar}{\ensuremath{\overline{\nu}_{e}}}                   
\newcommand{\enurec}{\ensuremath{E_\nu^\mathrm{rec}}}     
\newcommand{\deltacp}{\ensuremath{\delta_{CP}} }

\section{The Hyper-Kamiokande project}
Hyper-Kamiokande (HK) is a next generation underground water Cherenkov detector.
Based on the highly successful Super-Kamiokande (SK), the detector performance will be further enhanced by an order of magnitude larger fiducial mass and higher performance photodetectors. 
It will have far-reaching sensitivities for a very broad range of science topics, including neutrino oscillation studies, proton decay searches, and neutrino astrophysics.

\begin{figure}[b]
\centering
\includegraphics[width=0.52\textwidth]{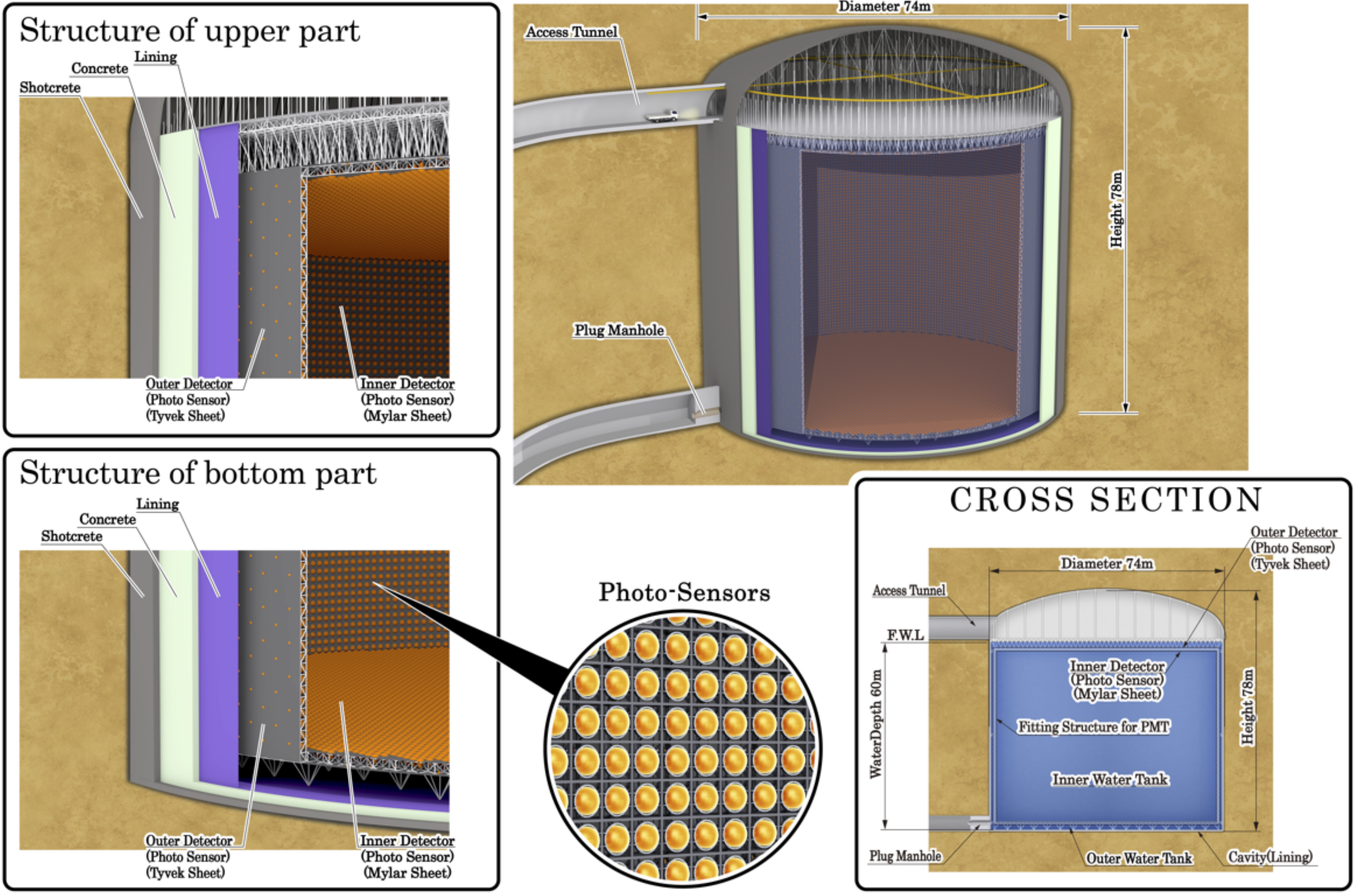}
\caption{Schematic view of one HK detector. From \cite{Hyper-Kamiokande:2016dsw}.}
\label{fig:HK-schematics}
\end{figure}

The current baseline design of HK comprises two cylindrical detectors that are 60~m in height and 74~m in diameter  (Fig.~\ref{fig:HK-schematics}).
The design was revised in the beginning of 2016 as a result of optimization, taking into account the recent technical development~\cite{Hyper-Kamiokande:2016dsw}.
The newly developed 50cm PMT, Hamamatsu R12860, has twice better photon detection efficiency and timing resolution compared to R3600, the PMT used for SK~\cite{jost}.
In addition, it has an improved pressure tolerance so that a deeper tank becomes feasible.
Alternative solutions for photosensors are also extensively studied by the international collaboration.

The total (fiducial) mass of water will be 260(190)~kton per tank. 
The inner detector region will be instrumented with 40,000 50~cm high-performance PMTs, corresponding to $40\%$ photo-cathode coverage.
The outer detector will be equipped with 6,700 20~cm PMTs.
The proposed location for HK is about 8~km south of SK and 1,750 meters water equivalent 
(or 648~m of rock) deep.
A staging between the first and second tank is planned for the construction.
In the study described in this paper, the second tank is assumed to become operational at the same site after six years.
Recently, a possibility of building the second tank in Korea is explored~\cite{Abe:2016ero}.
It is discussed separately~\cite{Korea}.

In this paper, the overview of HK physics program will be described, with an emphasis on the enhancement of capabilities realized with the new design.

\section{Physics capabilities}

\subsection{Search for proton decay}

\subsubsection{$p\to e^+ \pi^0$}
Proton decays into a positron and neutral pion, $p\to e^+ \pi^0$, are a dominant decay mode in many GUT models.
It also has a very clean experimental signature in a water Cherenkov detector with full reconstruction of the event.

After decades of search, the sensitivity is still improving with advancement of detector technology and analysis technique.
One of examples for such a technique is the background suppression with the neutron tagging.
In the proton decay events, the probability of neutron emission is rather small, while in the atmospheric neutrino events, which is the dominant background of proton decay searches, often neutrons are produced.
Thus, neutron tagging can provide an additional handle to suppress the background for the proton decay search and improve the sensitivity.

\begin{figure}[tb]
\centering
\includegraphics[width=0.95\textwidth]{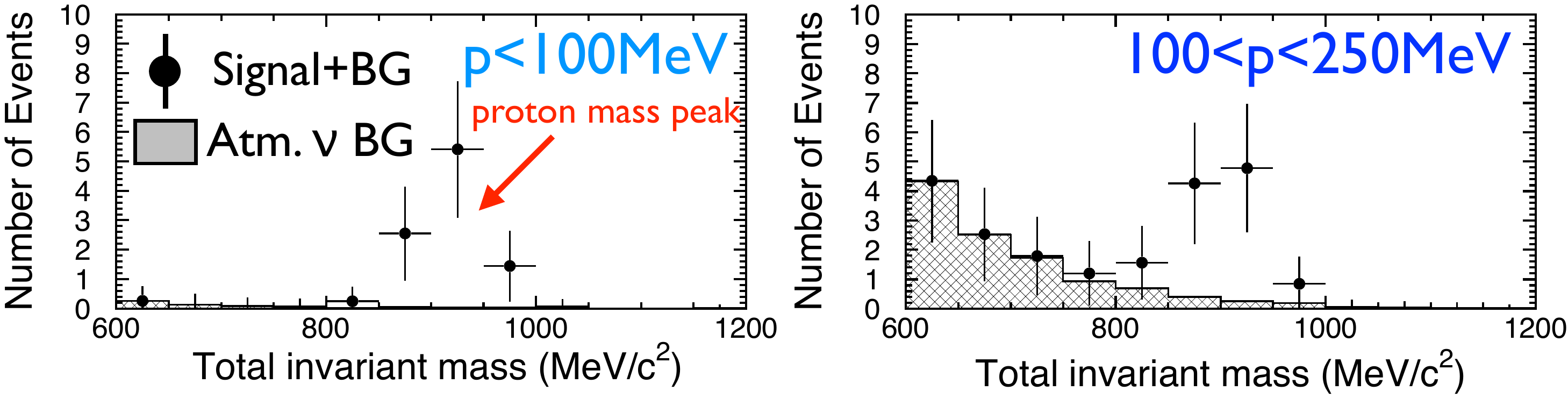}
\caption{Expected invariant mass distributions for $p \rightarrow e^{+} \pi^{0}$ candidates with ten years of HK.
The proton lifetime is assumed to be $1.7 \times 10^{34}$~years.
The left (right) plot shows the free (bound) proton enhanced region in the total momentum $p_\mathrm{tot}$, $p_\mathrm{tot}<100$~MeV ($100<p_\mathrm{tot}<250$~MeV).
The points (histogram) show the sum of the background and proton decay signal (atmospheric neutrino background).
}
\label{fig:pdecay}
\end{figure}

The ability to tag the 2.2\,MeV photon from neutron capture on hydrogen, $n+p \to d+\gamma$, in a water Cherenkov detector is demonstrated by the proton decay searches with SK-IV~\cite{Miura:2016krn}.
With 40\% photo-coverage and R3600 PMT, the neutron tagging efficiency in SK is about 20\%.
HK will have a better efficiency for the low energy photons thanks to the higher photon detection efficiency.
From the expected number of hits for 2.2\,MeV $\gamma$ ray,
the neutron tagging efficiency in HK is assumed to be 70\%.
Gd doping is also investigated as an option to further improve the neutron tagging efficiency, but not considered in this study.
Table~\ref{tbl:pdkepi0} shows the expected signal efficiency and background rates for HK,
compared to those of the SK-IV~\cite{Miura:2016krn}.
Figure~\ref{fig:pdecay} shows the reconstructed invariant mass distribution expected with ten years of HK data for the proton lifetime of $1.7 \times 10^{34}$~years.


\begin{table}[tb]
 \centering
  \caption{Signal efficiency and background rates 
           for  $p \rightarrow e^{+} \pi^{0}$ at HK and SK-IV~\cite{Miura:2016krn} .  }
  \label{tbl:pdkepi0}
  \begin{tabular}{l|cccc|cccc}
\hline\hline
        & \multicolumn{4}{c|}{ $0 < p_\mathrm{tot} < 100 \mbox{MeV/c}$ } &  \multicolumn{4}{c}{ $100 < p_\mathrm{tot} < 250 \mbox{MeV/c}$ } \\
        & $\epsilon_\mathrm{sig}$&$\sigma_{\epsilon}$ & Bkg & $\sigma_\mathrm{Bkg}$ &$\epsilon_\mathrm{sig}$ & $\sigma_{\epsilon}$& Bkg & $\sigma_\mathrm{Bkg}$\\
   &  [\%] &   [\%] &  [/Mton$\cdot$yr] &  [\%]   &   [\%]   &   [\%]     &  [/Mton$\cdot$yr] &  [\%] \\
\hline
HK & 18.7   & 6.5    &  0.06   &  32.8                 &     19.4   & 14.9  &  0.62     & 31.9                 \\
SK-IV & 18.7 & 10.2 & 0.18 & 32.7 & 19.4 & 17.7 & 1.1 & 31.1 \\
\hline
\hline
  \end{tabular}
\end{table}

\subsubsection{$p\to \bar{\nu} K^+$}
Proton decays into an antineutrino and a charged kaon, $p\to \bar{\nu} K^+$, are a dominant mode in many of supersymmetric grand unified theories.
In a water Cherenkov detector, $K^+$ from proton decay is not directly visible because its momentum is below the Cherenkov threshold,
but can be identified from its decay products. 

For $K^+ \to \mu^{+}+\nu$ (branching fraction 64\%), in addition to detecting a monochromatic (236\,MeV/$c$) muon,
nuclear de-excitation $\gamma$ ray (6.3\,MeV) can be used to tag the signal.
Better detection efficiency and timing resolution of the new photosensor will lead to an improved efficiency of low energy $\gamma$ rays.

For $K^+ \to \pi^{+}+\pi^{0}$ (branching fraction 21\%), 
the $\pi^+$ has a momentum just above the Cherenkov threshold and emits only faint light.
With better photon detection efficiency, the detection efficiency of $\pi^+$ will be improved.

Table~\ref{tbl:pdknuk} summarizes the efficiency and background expectation for $p\to \bar{\nu} K^+$ searches.
Neutron tagging is also applied for this mode.
There is the third method for the $p\to \bar{\nu} K^+$ search, search for an excess in the muon momentum distribution, not shown in the table.

\begin{table}[tb]
 \centering
  \caption{Signal efficiency and background rates  for $p \rightarrow \bar{\nu} K^{+}$. The numbers for SK~\cite{Abe:2014mwa} are also listed.}
  \label{tbl:pdknuk}
  \begin{tabular}{l|cccc|cccc}
\hline\hline
  & \multicolumn{4}{c|}{ $K^+ \to \mu^{+}+\nu$ with prompt $\gamma$ } &  \multicolumn{4}{c}{ $K^+ \to \pi^{+}\pi^{0}$  }  \\
  &  $\epsilon_\mathrm{sig}$ & $\sigma_{\epsilon}$ & Bkg & $\sigma_\mathrm{Bkg}$& $\epsilon_\mathrm{sig}$ & $\sigma_{\epsilon}$ & Bkg & $\sigma_\mathrm{Bkg}$  \\
  &[\%] &   [\%] &  [/Mton$\cdot$yr] &  [\%] &  [\%] &   [\%] & [/Mton$\cdot$yr]  & [\%]  \\
\hline
HK & 12.7  & 19.0    &  0.9   &  27.0  &  10.8   & 10.0  &  0.7     & 31.0              \\
SK-IV & 9.1 &  22 & 1.5 & 25 & 10.0 & 9.5 & 2.0 & 29 \\
\hline
\hline
  \end{tabular}
\end{table}

\begin{figure}[tb]
\centering
\includegraphics[width=0.9\textwidth]{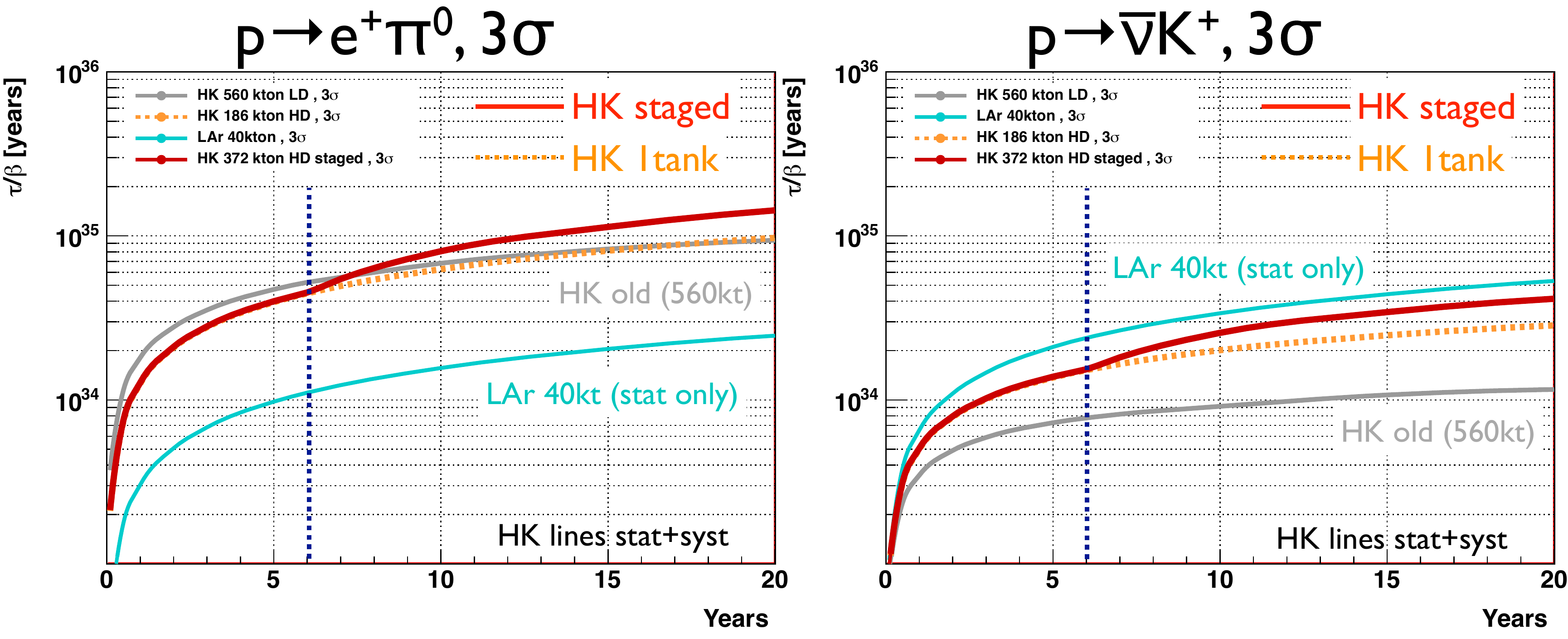}
\caption{Estimated 3~$\sigma$ discovery potential of $p \to e^+ \pi^0$ (left) and $p \rightarrow \bar \nu K^{+}$ (right) as a function of run time.
The red, orange and grey lines correspond to the baseline design of HK, the case for just single tank, and the old design (560~kt fiducial mass with 20\% coverage with R3600), respectively.
The cyan line shows a 40~kton liquid argon detector, with assumption of a signal efficiency of 45\%(97\%) and 
background of 1.0(1.0) events per~Megaton$\cdot$year for $p \to e^+ \pi^0$ ($p \rightarrow \bar \nu K^{+}$).
Systematic errors are included for the HK lines but not for the liquid argon detector.
}
\label{fig:PdecaySensitivity}
\end{figure}

\subsubsection{Expected sensitivity for proton decay}
Figure~\ref{fig:PdecaySensitivity} shows the estimated 3~$\sigma$ discovery potential of $p \to e^+ \pi^0$ (left) and $p \rightarrow \bar \nu K^{+}$ (right) as a function of run time.
After 20 years of operation, the expected 3~$\sigma$ sensitivities for $p \to e^+ \pi^0$ and $p \rightarrow \bar \nu K^{+}$ are 8.0$\times 10^{34}$ and 2.5$\times 10^{34}$ years, respectively.
With HK, there is a large potential to discover the proton decays with lifetime beyond the current lower limit given by SK, 1.6$\times 10^{34}$ years and 5.9$\times 10^{33}$ years.

\subsection{Low energy physics}
As is demonstrated by SK, a water Cherenkov detector has an excellent potential for broad and rich science with low energy ($\mathcal{O}(1$--$10)$~MeV) neutrinos.
In this energy region, the performance of the detector is mainly limited by number of detected photons.
Hence, improved photon detection efficiency of HK will significantly benefit the low energy physics program.
In order to fully exploit the improved performance for low energy physics, 
careful control of background (such as radioactivity and spallation) and water quality, and precise calibration by design and operation of the detector will be necessary.
An extensive R\&D  by international cooperation is under way.

One example of low energy physics that benefits from HK design is the study of solar neutrinos.
There is about 2\,$\sigma$ of tension in $\Delta m^2_{21}$
between solar neutrino measurements and 
KamLAND reactor neutrino measurement~\cite{Abe:2016nxk}.
The day-night asymmetry of solar neutrino event rate
due to the regeneration of the electron neutrinos through the MSW
matter effect in the Earth can provide us a precise determination of $\Delta m^2_{21}$
using $\nue$ (in contrast to $\nuebar$ from reactors),
providing a 5~$\sigma$ resolution if the difference from KamLAND value persists and 0.3\% systematic uncertainty is achieved.
The observation of the spectrum upturn in solar neutrino,
caused by the transition of the survival probability from the matter
dominant to the vacuum dominant energy region, will be possible with the low threshold and high statistics.
The precise measurement of the spectrum shape can distinguish the standard neutrino
oscillation scenario from exotic models.
5~$\sigma$ observation of the upturn in the transition region will be possible with HK, if background and calibration level similar to those of SK can be achieved.
Also, thanks to the good resolution, measurement of \textit{hep} neutrino could be possible.

Another important topic is the study of supernova neutrinos, as presented in~\cite{jost}.

\subsection{Studies of neutrino oscillation}
\subsubsection{Long baseline program}
Recently, the T2K collaboration reported the first
constraint on $\deltacp$~\cite{Abe:2015awa, Abe:2017uxa, T2K}, 
which indicates that the $CP$ violation in the lepton sector
may be large, although the statistical significance is still insufficient.
The observation and study of the $CP$ asymmetry in the lepton sector,
now possible with the comparison of $\numu \to \nue$ and $\numubar \to \nuebar$ oscillations,
is one of the most important topics in particle physics.
Precision measurements of oscillation parameters require both large statistics
and well controlled systematics.  Combining an intense and
high quality neutrino beam from J-PARC, the huge mass and high
performance of Hyper-K detector, a highly capable near/intermediate
detector complex~\cite{ND}, and the full expertise obtained from ongoing T2K/SK
experiments, Hyper-K will be the best project to probe the $CP$
violation in the lepton sector and new physics with neutrino oscillation.

The beam power from J-PARC accelerator and its neutrino beamline is expected to be significantly increased in near future. 
The accelerator upgrade to double the repetition rate is ongoing, 
to reach the design power of 750\,kW and beyond.
Based on high intensity studies of the current accelerator performance,
it is expected that 1.3\,MW beam power can be achieved by the time HK will start operation.
The upgrade of the neutrino beamline is also planned to keep up with the upgrade of accelerator power.
For the sensitivity estimation of the long baseline experiment using HK, an integrated beam power of 1.3~MW$\times 10^8$~sec, corresponding to ten years, is assumed.

\begin{table}[tbp]%
\caption{\label{Tab:sens-selection-nue}%
The expected number of $\nue/\nuebar$ candidate events and
efficiencies with respect to FCFV events.
Normal mass hierarchy with
$\sin^22\theta_{13}=0.1$ and $\deltacp=0$ are assumed.  Background is
categorized by the flavor before oscillation.}
\begin{center} 
{\scriptsize
\begin{tabular}{cc|cc|ccccc|c|c} \hline \hline
&	& \multicolumn{2}{c|}{signal} & \multicolumn{6}{c|}{BG} & \multirow{2}{*}{Total} \\ 
&	&$\numu \to \nue$	& $\numubar \to \nuebar$ 	&$\numu$ CC &$\numubar$ CC	&$\nue$  CC& $\nuebar$ CC & NC & BG Total	&  \\ \hline  
\multirow{2}{*}{$\nu$ mode} & Events	& 2300	&	21& 10 	& 0	 & 347	&15	& 188	&	560 & 2880 \\ 
 & Eff.(\%)  & 63.6 & 47.3 & 0.1 & 0.0 & 24.5 & 12.6 & 1.4 & 1.6 & --- \\
 \hline
\multirow{2}{*}{$\bar{\nu}$ mode} & Events	& 289	&	1656	& 3	& 3	& 142	& 302	& 274&	724 & 2669 \\ 
 & Eff. (\%) & 45.0 & 70.8 & 0.03 & 0.02 & 13.5 & 30.8 & 1.6 & 1.6 & --- \\
\hline \hline
\end{tabular}%
}
\end{center}
\end{table}%

Table~\ref{Tab:sens-selection-nue} shows the expected number of events after $\nu_e$ signal selection, for $\sin^22\theta_{13} =0.1$, $\delta=0$, and normal mass hierarchy.
For each of neutrino and anti-neutrino mode, $\mathcal{O}(1000)$ signal events are expected.
The sensitivity is estimated based on a framework developed in T2K experiment~\cite{Abe:2014tzr}.
The analysis is the same as one described in \cite{Abe:2015zbg}, except for the update of the systematic uncertainty estimate.
A binned likelihood analysis based on the reconstructed neutrino energy
distribution is performed using both \nue\ (\nuebar) appearance
and \numu\ (\numubar) disappearance samples simultaneously.
The systematic uncertainty is estimated based on the experience of T2K, with an extrapolation considering improvement expected in HK era.
Correlations of systematics between energy bins and flavors are taken into account using an error matrix.

\begin{figure}[tb]
\centering
\includegraphics[width=0.48\textwidth]{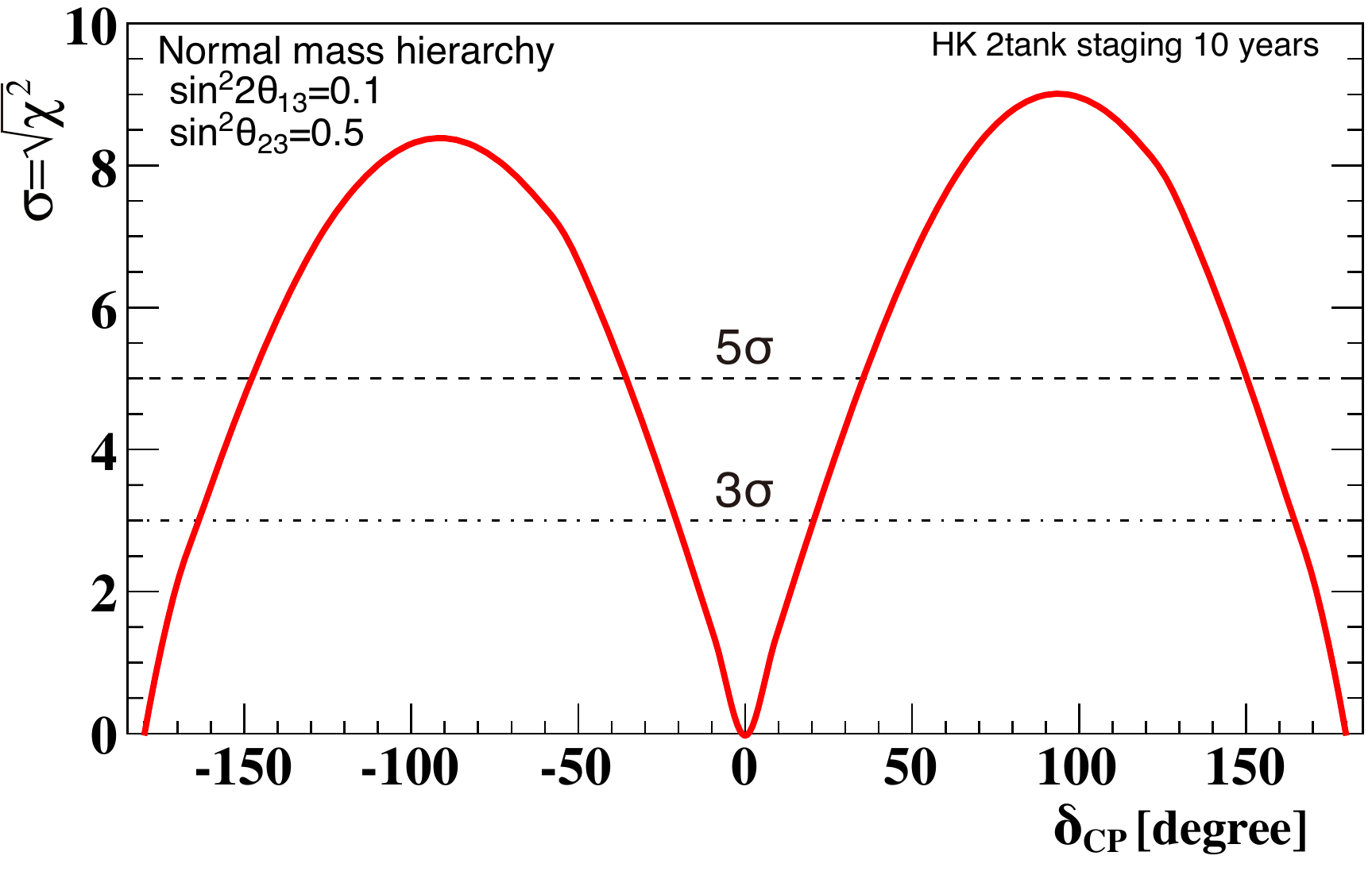}
\includegraphics[width=0.48\textwidth]{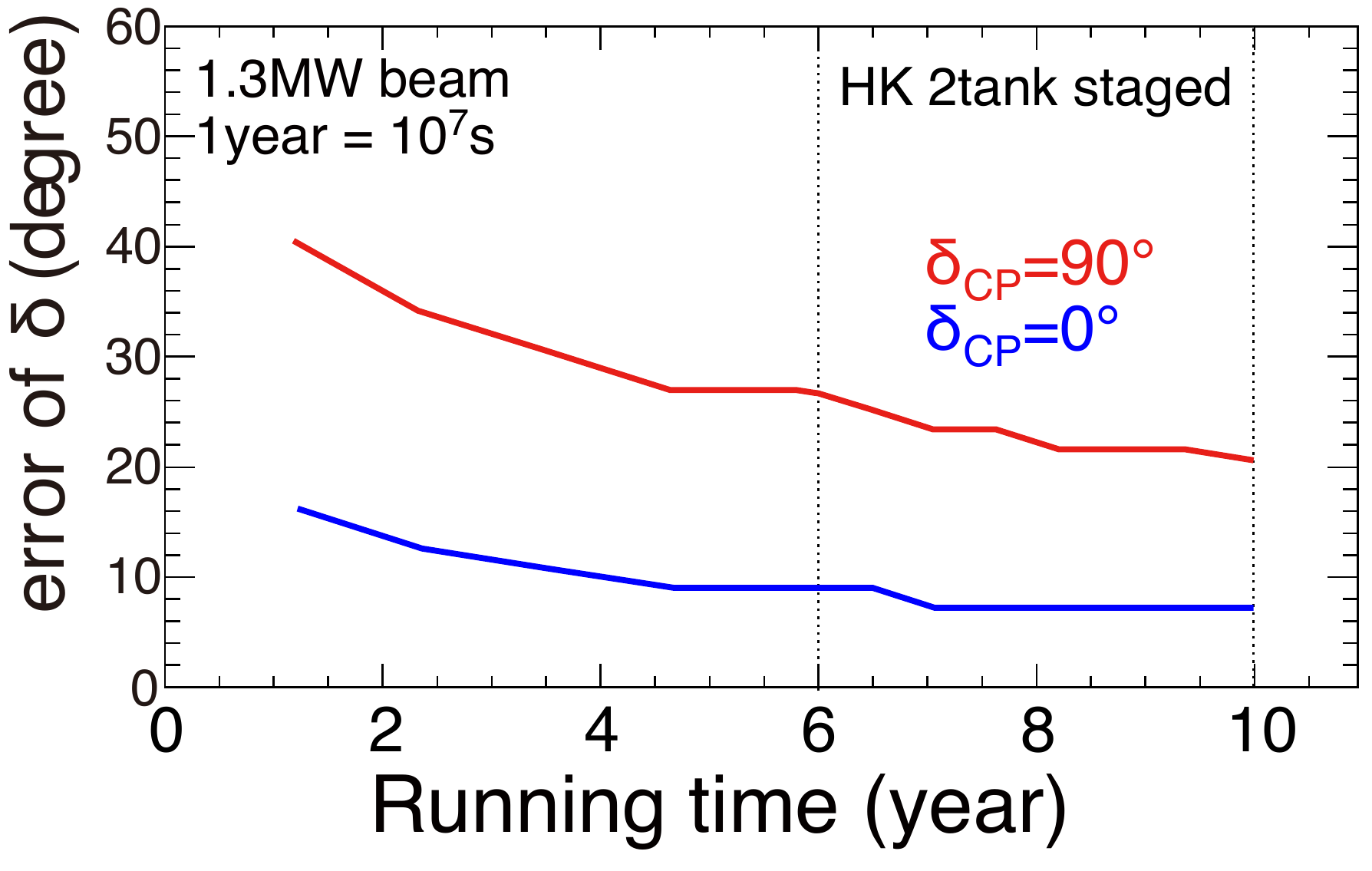}
\caption{(Left) Expected significance to exclude $\sin\deltacp = 0$  for normal mass hierarchy. (Right) Expected 68\% CL uncertainty of $\deltacp$ as a function of running time.}
\label{fig:CPsens}
\end{figure}

Figure~\ref{fig:CPsens}(left) shows the expected significance to exclude
$\sin\deltacp = 0$ (the $CP$ conserved case). 
$CP$ violation in the lepton sector can be observed with more than
3(5)\,$\sigma$ significance for 78(62)\% of the possible values of
$\deltacp$.
Figure~\ref{fig:CPsens}(right) shows the 68\% CL uncertainty of
$\deltacp$ as a function of the running time.
The value of $\deltacp$ can be determined with an
uncertainty of 7.2$^\circ$ for $\deltacp=0^\circ$ or $180^\circ$, and
21$^\circ$ for $\deltacp=\pm90^\circ$.

Using both $\nu_e$ appearance and $\nu_\mu$ disappearance channels,
precise measurements of $\sin^2\theta_{23}$ and $\Delta m^2_{32}$ will
be possible.
Expected 1$\sigma$ uncertainty of
$\sin^2\theta_{23}$ is 0.015(0.006) for $\sin^2\theta_{23}=0.5(0.45)$.
The uncertainty of $\Delta m^2_{32}$ is expected to reach $<1\%$.

There will be also a variety of measurements possible with both near
and far detectors, such as neutrino-nucleus interaction cross section
measurements and search for exotic physics, using the well-understood
neutrino beam.

\subsubsection{Atmospheric neutrinos}
Atmospheric neutrinos provide a wide variety of energy, baseline, and flavor of neutrinos,
giving access to complementary information to accelerator neutrinos.
In particular, significant modification of neutrino oscillation probabilities in the energy range 2-10 GeV due to matter effects inside the Earth gives a sensitivity to the mass hierarchy.
After 10 years, the measurement with atmospheric neutrino alone is expected to resolve the mass hierarchy at $\sqrt{\Delta \chi^{2}} > 3$ for both hierarchy assumptions and when $\sin^2\theta_{23} > 0.45$.

\begin{figure}[tb]
\centering
\includegraphics[width=0.9\textwidth]{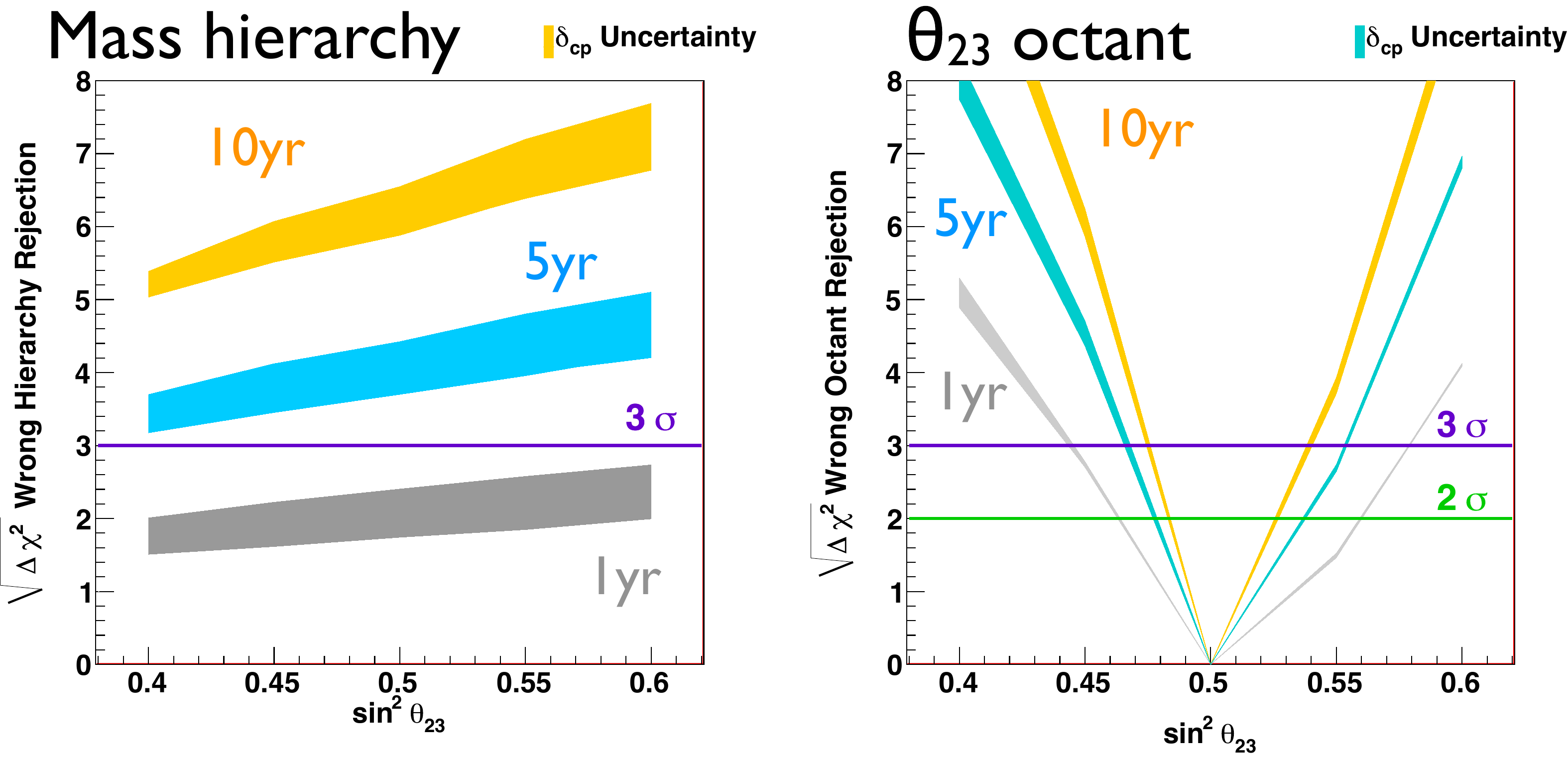}
\caption{Mass hierarchy (left) and $\theta_{23}$ octant (right) sensitivity by a combination of beam and atmospheric neutrinos in HK.}
\label{fig:atm}
\end{figure}

Moreover, the sensitivities are enhanced by a combination of accelerator and atmospheric neutrinos.
Figure~\ref{fig:atm} shows the sensitivities for mass hierarchy (left) and $\theta_{23}$ octant (right) by a combination of beam and atmospheric neutrinos.
The mass hierarchy can be determined with more than 3(5) $\sigma$ with five (ten) years of data, and the octant of $\theta_{23}$ can be resolved if $|\theta_{23} - 45| > 2.5^{\circ}$  in ten years.

Atmospheric neutrino can study additional topics, such as $\nu_\tau$ cross section measurement, search for sterile neutrino, and the test of Lorentz invariance.
It will also provide information on the chemical composition of Earth's outer core using matter effect, contributing to the geophysics.

\section{Conclusions}
Hyper-Kamiokande will have a rich program with world-leading science output.
Based on technology well established with past/ongoing experiments,
it will realize a fast and robust approach to the $CP$ violation in the lepton sector and a long term observational program with a wide range of science.
An international collaboration of about 300 members from 15 countries is working to promote the project towards realization.
Assuming the start of construction in 2018, the data taking is expected to start from 2026.

\Acknowledgements
The work was partially supported by JSPS Grant-in-Aid for Scientific Research on Innovative Areas ``Unification and Development of the Neutrino Science Frontier'' A03 (25105004), and JSPS/RFBR under Japan-Russia Research Cooperative Program.

\end{document}